# Ranking forestry journals using the *h*-index

*Jerome K Vanclay*
*Southern Cross University*
*PO Box 157, Lismore NSW 2480, Australia*
*JVanclay@scu.edu.au*

## Abstract

An expert ranking of forestry journals was compared with journal impact factors and *h*-indices computed from the ISI Web of Science and internet-based data. Citations reported by Google Scholar offer an efficient way to rank all journals objectively, in a manner consistent with other indicators. This *h*-index exhibited a high correlation with the journal impact factor (r=0.92), but is not confined to journals selected by any particular commercial provider. A ranking of 180 forestry journals is presented, on the basis of this index.

## Introduction

The Thomson Scientific (TS) Journal Impact Factor (JIF; Garfield, 1955) has been the dominant measure of journal impact, and is often used to rank journals and gauge relative importance, despite several recognised limitations (Hecht et al., 1998; Moed et al., 1999; van Leeuwen et al., 1999; Saha et al., 2003; Dong et al., 2005; Moed, 2005; Dellavalle et al., 2007). Other providers offer alternative journal rankings (e.g., Lim et al., 2007), but most deal with a small subset of the literature in any discipline. Hirsch's *h*-index (Hirsch, 2005; van Raan, 2006; Bornmann & Daniel, 2007a) has been suggested as an alternative that is reliable, robust and easily computed (Braun et al., 2006; Chapron and Husté, 2006; Olden, 2007; Rousseau, 2007; Schubert and Glänzel, 2007; Vanclay, 2007; 2008). The *h*-index has also been used to rank researchers (Oppenheim, 2006; Bornmann and Daniel, 2007b; Grant et al., 2007; Schreiber, 2007), institutions (Prathap, 2006; Bar-Ilan, 2007; Smith, 2008) and topics. This study presents an analysis of the JIF, *h*-index, and other indicators of journal utility, with a view to ranking forestry literature.

In preparation for the Australian government's Research Quality Framework (RQF; Gale et al., 2005; DEST, 2007), professional bodies in Australia were asked to identify and rank relevant journals within their discipline into four prestige bands, based on journal quality. Participants were asked to allocate journals to one of four classes, representing the top 5 percentile (A1), the 80-95 percentile (A), the 50-80 percentile (B), and the residue (C). The classification offered by the Institute of Foresters of Australia (*pers comm.*, 21 November 2007) implied a ranking substantially different to the JIF, even though the 2005 JIF data were available to members to assist them in their classification. The wide range of JIFs within an assigned band was noteworthy, as was the disagreement regarding the top journal. This study attempts to shed some light on this discrepancy.

## Method

The study draws on subjective journal rankings proposed by four individuals, nominated by and senior members of the Institute of Foresters of Australia, which was asked by the Australian Academy of Technological Sciences and Engineering (ATSE, 2007) to assist in ranking forestry journals in terms of academic standing. The author played no part in the selection of these experts, and the ranking offered by the author has been omitted from this analysis. Three of the experts had a PhD, and represented current or past heads of a university department, a national research agency, a development assistance agency, and a consultancy firm.

The Institute of Foresters of Australia publishes one of the journals under consideration, *Australian Forestry*. Three of the four experts placed *Australian Forestry* in the top 15% of journals, whereas this study suggests that it is near the 76 percentile, suggesting some parochial bias by the experts. However, the rankings by the individual experts tended to be consistent, exhibiting correlations of $r \geq 0.69$ (Figure 1; Table 1).

This study also draws on Journal Impact Factors from the 2006 Journal Citation Reports, and on *h*-indices computed automatically from two sources, the ISI Web of Science (Thomson Scientific, version 4.0, WoS) and Harzing's (2007) Publish or Perish (PoP), a software package that harvests data from Google Scholar (GS), a specialised internet search engine restricted to scholarly documents (Noruzi, 2005; Pauly & Stergiou, 2005; Meho & Yang, 2007; Kousha & Thelwall, 2008).

Although the *h*-index is robust (Vanclay 2007), automated calculation may be biased by typographic and other database errors (Jacso 2008). Several precautions were adopted minimize such bias. The *h*-index calculation was performed both using the full journal title and using common abbreviations (e.g., to detect problems such as *Ann. Forest Sci.* which is not recognised by GS as *Annals of Forest Science*). Citation lists reported by PoP were sorted by author and by title to facilitate detection and correction of typographic errors and missing details (e.g., such as the lack of machine-readable publication dates in *Tree-Ring Research*).

Hirsch's h-indices were computed for several intervals (Table 1), but the 8-year interval 2000-2007 seemed insightful for forestry journals, many of which have a long cited half-life. The *h*-indices computed from WoS and GS data are similar ($r = 0.93$, $n = 43$ for 2000-2007 data), but the former are available only for WoS-listed journals (about 15% of forestry journals), whereas the latter can be computed for any journal or citation visible to Google Scholar.

## Results

Table 1 and Figure 1 illustrate the correspondence between a classification allocated by experts and the JIF, for each of the four contributors and the 27 journals recognised by both ATSE (2007) and WoS. There was a considerable discrepancy between the assigned classification and the JIF-based ranking of forestry journals. In Figure 1, the spread of points and the weak trend illustrate the magnitude of the differences between experts and the ranking implied by the JIF. The shape of the trend is not unexpected, because the WoS data are censored to represent the top few journals (about 15%). Although variants of *h*-index is well correlated with the JIF ($r \geq 0.75$; Table 1), it exhibits closer agreement with the expert assessment ($r \geq 0.52$) than does the JIF ($r=0.52$), suggesting that the *h*-index may be useful for ranking journals objectively. An advantage of the PoP *h*-index is that it may be computed for the many journals not acknowledged by Thomson Scientific.

**Table 1**. Journal impact factors contrasted with an expert classification of 27 forestry journals by four individuals into four classes.

| Journal | Expert assignment | | | | Weighted score† | ISI JIF | ISI *h*-index | PoP *h*-index 2000-07 | PoP lifetime ‡ *h*-index |
|---|---|---|---|---|---|---|---|---|---|
| | A1 | A | B | C | | | | | |
| Forest Ecology and Management | 4 | | | | 3.90 | 1.8 | 36 | 43 | 69 |
| Agricultural and Forest Meteorology | 2 | 2 | | | 3.75 | 2.9 | 43 | 41 | 67 |
| Tree Physiology | 1 | 3 | | | 3.68 | 2.3 | 35 | 28 | 41 |
| Annals of Forest Science | 3 | | 1 | | 3.60 | 1.3 | 18 | 19 | 32 |
| International Forestry Review | | 4 | | | 3.60 | 0.6 | 8 | 12 | 18 |
| Forestry | | 4 | | | 3.60 | 0.8 | 14 | 16 | 31 |
| Aust J Bot | | 4 | | | 3.60 | 0.9 | 30 | 21 | 40 |
| Trees-Structure and Function | | 4 | | | 3.60 | 1.5 | 20 | 22 | 36 |
| Canadian Journal of Forest Research | | 4 | | | 3.60 | 1.5 | 33 | 23 | 18 |
| New Forests | | 3 | 1 | | 3.38 | 0.7 | 10 | 11 | 25 |
| Silva Fennica | | 3 | 1 | | 3.38 | 0.9 | 17 | 14 | 23 |
| International Journal of Wildland Fire | | 3 | 1 | | 3.38 | 1.7 | 17 | 21 | 29 |
| Forest Science | 2 | | 2 | | 3.30 | 1.5 | 20 | 19 | 54 |
| Silvae Genetica | | 2 | 2 | | 3.15 | 0.3 | 8 | 9 | 23 |
| Forest Policy and Economics | 1 | | 3 | | 3.00 | 0.9 | 11 | 17 | 16 |
| Journal of Forestry | | 1 | 3 | | 2.93 | 1.2 | 18 | 8 | 37 |
| European Journal of Forest Research | | | 4 | | 2.70 | 0.8 | 7 | 6 | 22 |
| Forest Pathology | | | 4 | | 2.70 | 0.7 | 12 | 11 | 11 |
| Wood Science & Technology | | | 4 | | 2.70 | 0.7 | 12 | 13 | 30 |
| Forest Products Journal | | | 4 | | 2.70 | 0.4 | 13 | 14 | 19 |
| Scandinavian Journal of Forest Research | | | 4 | | 2.70 | 0.9 | 16 | 18 | 25 |
| Journal of tropical forest science | | | 3 | 1 | 2.28 | 0.2 | 3 | 7 | 15 |
| Forestry Chronicle | | | 3 | 1 | 2.28 | 0.8 | 14 | 13 | 20 |
| Agroforestry Systems | | | 3 | 1 | 2.28 | 0.9 | 15 | 19 | 39 |
| Northern J Applied Forestry | | | | 4 | 1.00 | 0.8 | 6 | 6 | 11 |
| Western Journal of Applied Forestry | | | | 4 | 1.00 | 0.5 | 6 | 8 | 17 |
| Southern Journal of Applied Forestry | | | | 4 | 1.00 | 0.7 | 6 | 9 | 21 |
| **Correlations** | | | | | | | | | |
| Aggregate score | | | | | 1 | 0.52 | 0.64 | 0.61 | 0.52 |
| ISI JIF | | | | | 0.52 | 1 | 0.88 | 0.84 | 0.75 |
| ISI *h*-index | | | | | 0.64 | 0.88 | 1 | 0.90 | 0.76 |
| 2000-7 PoP *h*-index | | | | | 0.61 | 0.84 | 0.90 | 1 | 0.82 |
| lifetime PoP *h*-index | | | | | 0.52 | 0.75 | 0.76 | 0.82 | 1 |

† Score computed with weights 0.975, 0.9, 0.675 and 0.25 reflecting the percentile represented by A1, A, B and C (95-100, 85-95, 50-85 and 0-50% respectively).

‡ 'Lifetime' implies unconstrained by date, drawing on all entries within the database.

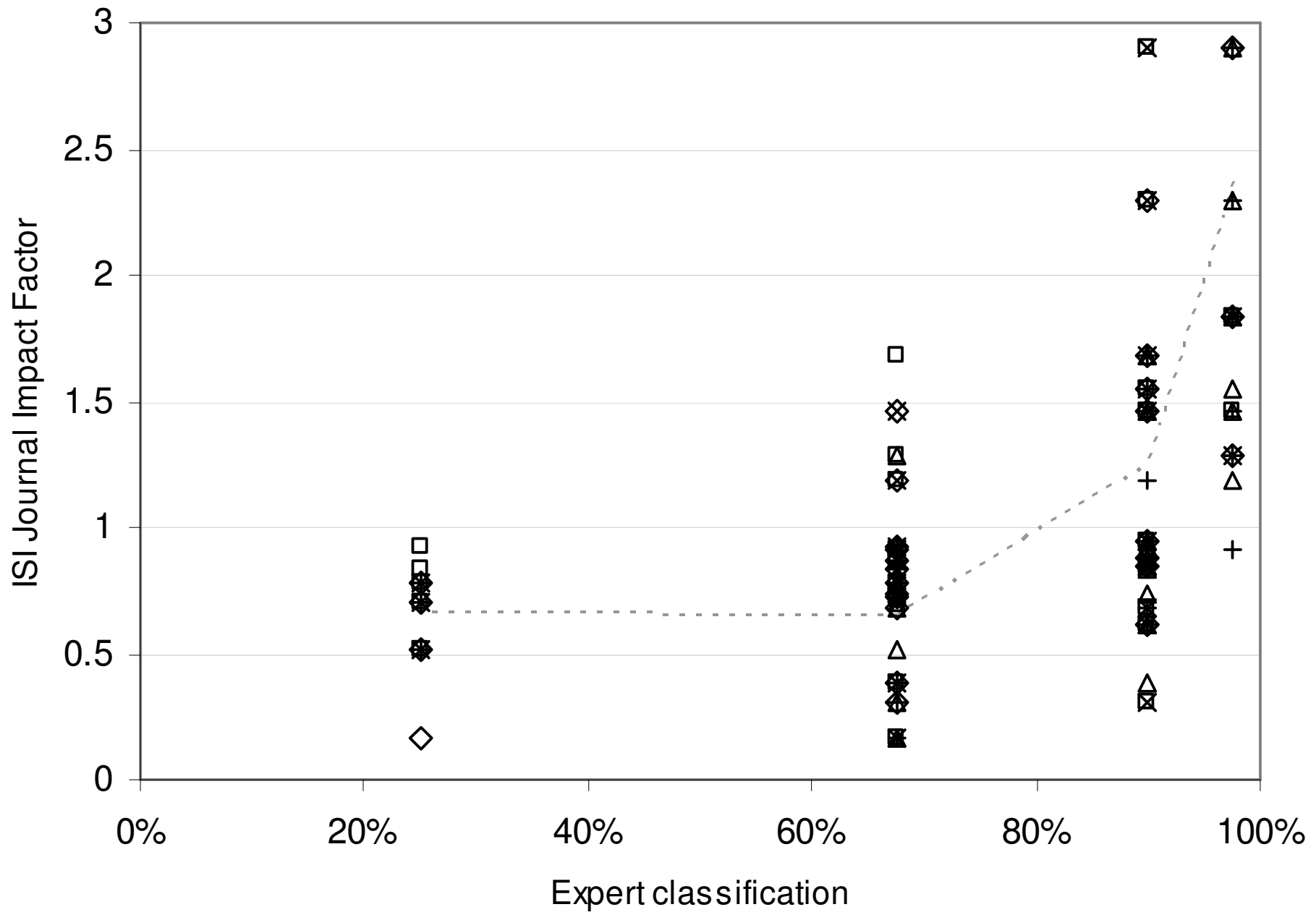

**Figure 1**. Journal impact factors contrasted with an expert classification of 27 forestry journals by four individuals into four classes (using different symbols for each expert).

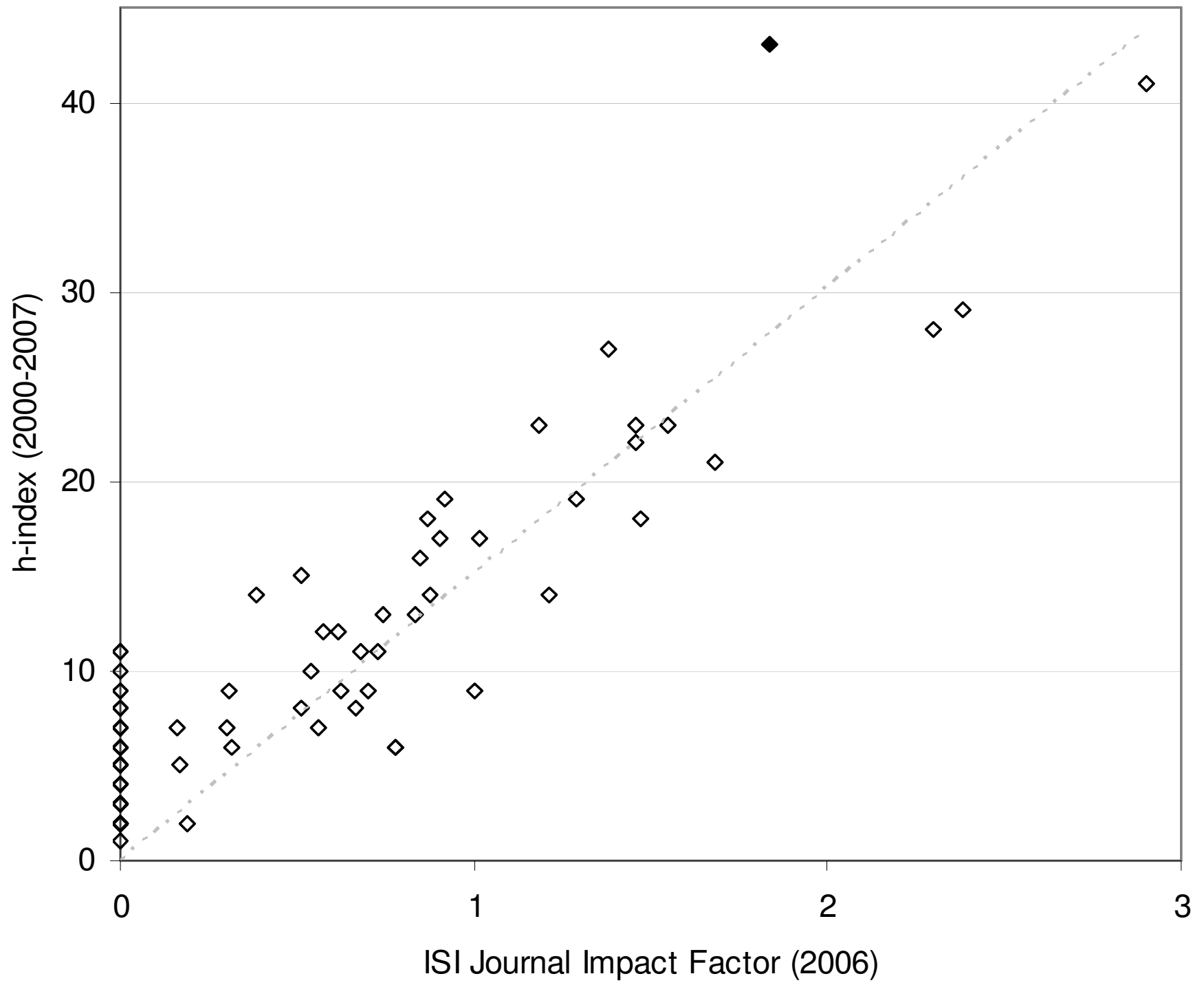

**Figure 2**. The relationship between the JIF and the PoP *h*-index (based on all citations accruing to journal publications during 2000-2007). The filled point near the top of the figure is *Forest Ecology and Management; Agricultural and Forest Meteorology* is at the top right. Journals not recognised by Thomson Scientific are shown with a zero JIF, and are omitted from the calculation of the trend line (trend based on 43 journals).

Expert ranking of two journals, *Agricultural and Forest Meteorology* (AFM) and *Forest Ecology and Management* (FEM), differed greatly to that implied by the JIF. The former has a higher JIF, but experts ranked the latter as more influential, as did the *h*-index (Table 1, Figure 2). Table 2 lists some key differences between these journals: AFM has a relatively small number of contributions, many of which are cited soon after publication, whereas FEM has a higher volume and is slower to accrue citations. Overall, the *h*-indices of the journals are comparable, but there is a tendency for WoS to report higher statistics for AFM, and for PoP to report higher statistics for FEM. Superficial examination of Table 1 may lead to the suggestion that AFM publishes relatively few papers all of which are high-quality, reflecting a high editorial standard, and in turn, credit to any author who has a paper accepted for publication (which is what the RQF seeks to achieve). However, this interpretation is simplistic, and warrants closer examination.

**Table 2**. Statistics for two of the top-ranked forestry journals.

| Indicator | *Agricultural and Forest Meteorology* | *Forest Ecology and Management* |
|---|---|---|
| Panel assessment | A1/A2 (95 percentile) | A1 (95-100 percentile) |
| Year established | 1964 | 1977 |
| JIF (2006) | 2.903 | 1.839 |
| Immediacy | 0.669 | 0.356 |
| Cited half-life | 6.7 | 5.8 |
| Total articles | 130 | 601 |
| Lifetime *h*-index (WoS) | 60 | 58 |
| *h*-index 2005-6 | 12 | 12 |
| *h*-index 2000-7 | 43 | 36 |
| Total cites 2000-7 | 9 113 | 21 470 |
| Lifetime *h*-index (PoP) | 67 | 69 |
| *h*-index 2005-6 | 9 | 12 |
| Mean cites/paper 2005-6 | 2.09 | 1.67 |
| *h*-index 2000-7 | 41 | 43 |
| Total cites 2000-7 | 8 544 | 25 913 |

The RQF seeks a proxy for research quality, and assumes that acceptance and publication by a journal indicates attainment of the standard indicated by the journal's ranking. The JIF is deficient for this purpose, because it reflects the average number of citations, and may conceal many 'free-riders' (Walter et al., 2003). Table 3 examines this issue, year-by-year for the last decade, and tabulates the proportion of papers in each journal that remain uncited (Weale et al., 2004), or fail to accrue at least one citation per year since publication. Despite its lower JIF, FEM has a lower proportion of papers that remain uncited, or that remain infrequently cited, for almost every year during the past decade, suggesting that by these yardsticks, FEM may be the journal that reflects better on contributors. This conclusion from Table 3 is reflected in the *h*-index, but not in the JIF (Table 2). Table 3 also illustrates that the *h*-index appears to plateau after eight years (i.e., in 2000), at least for these two forestry journals.

**Table 3**. Annualised data for two forestry journals, using h-indices calculated at end 2007.

| Year | *Agricultural and Forest Meteorology* | | | | *Forest Ecology and Management* | | | |
|---|---|---|---|---|---|---|---|---|
| | *h*-index (WoS) | *h*-index (PoP) | Fraction uncited (annualized %) | Not cited >1/year (%) | *h*-index (WoS) | *h*-index (PoP) | Fraction uncited (annualized %) | Not cited >1/year (%) |
| 2007 | 3 | | | | 2 | | | |
| 2006 | 6 | 4 | 49 | 49 | 7 | 6 | 62 | 62 |
| 2005 | 12 | 9 | 45 | 39 | 12 | 13 | 46 | 38 |
| 2004 | 18 | 15 | 54 | 35 | 16 | 18 | 47 | 34 |
| 2003 | 19 | 17 | 65 | 48 | 21 | 24 | 53 | 33 |
| 2002 | 20 | 17 | 71 | 45 | 26 | 29 | 55 | 33 |
| 2001 | 22 | 20 | 67 | 42 | 24 | 30 | 64 | 38 |
| 2000 | 24 | 27 | 71 | 39 | 30 | 35 | 64 | 38 |
| 1999 | 24 | 25 | 65 | 34 | 28 | 34 | 71 | 40 |
| 1998 | 21 | 23 | 70 | 35 | 31 | 34 | 73 | 44 |
| Mean | | | 62 | 41 | | | 59 | 40 |

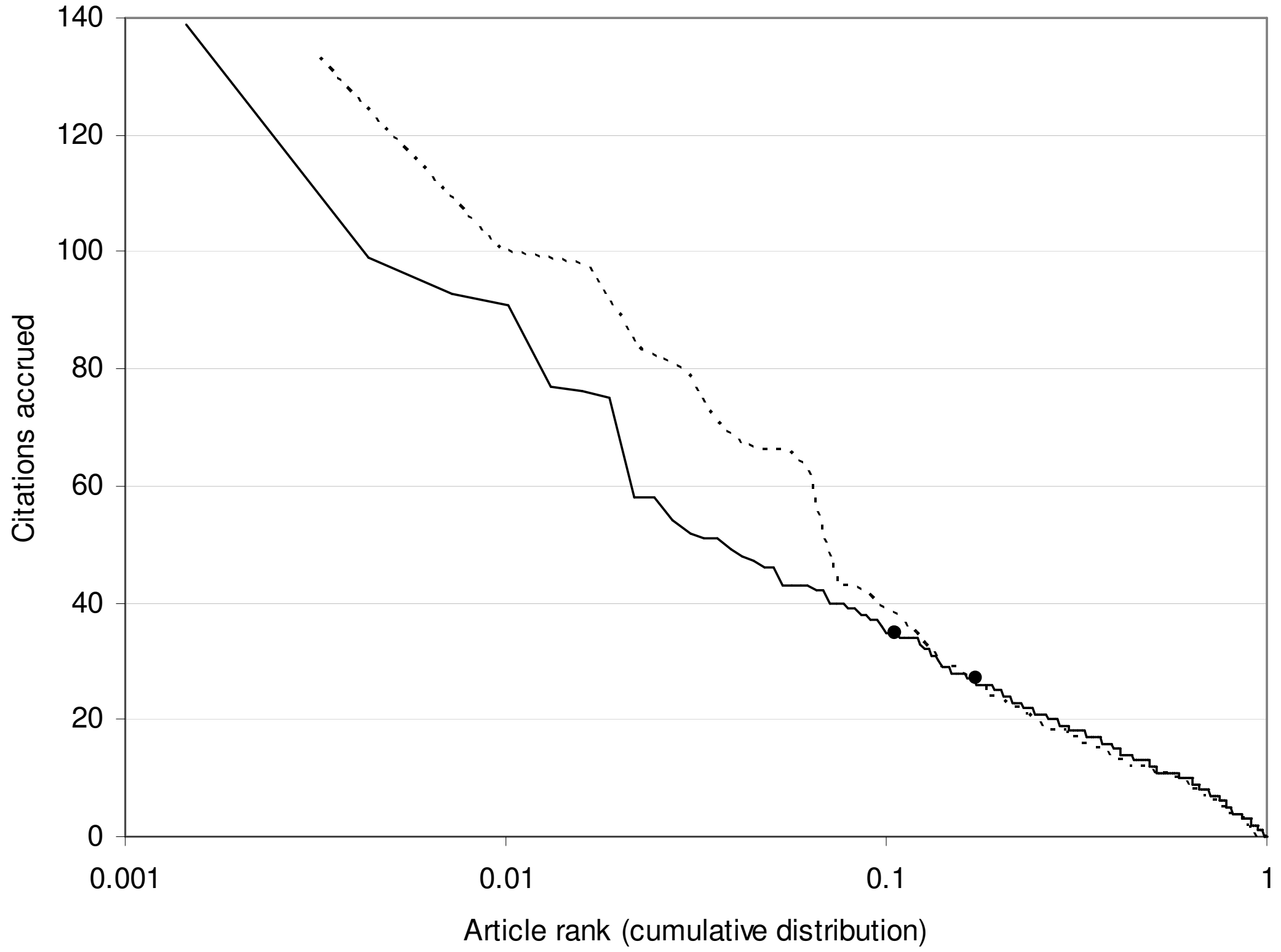

**Figure 3**. Pattern of citation accrual to two journals, *Agricultural and Forest Meteorology* (dotted) and *Forest Ecology and Management* (solid), using data from PoP. Note that the linear trend in the right-most part of the figure includes the point indicating the *h*-index (•).

Figure 3 illustrates the trends in citations to individual papers published in these two journals during the year 2000. The publication year 2000 was chosen because it reflects the half-life of these journals, and allows citation patterns to be fully expressed (Table 3; also Vanclay, 2008). Figure 3 reveals the number of citations for each paper in rank order, scaled to reflect the cumulative distribution function because of a three-fold difference in the number of papers published in these two journals. A logarithm scale is used because the great majority of papers accrue few citations, and exhibit a log-linear trend in their citation rate.

Figure 3 shows that the two journals have a very similar pattern of citation accrual to the majority of contributions, and that it is only in the most-frequently-cited 10% of papers that differences in citations appear. This equivalence is reflected in the *h*-indices (27 for AFM, 35 for FEM, PoP data), but not in the JIFs of the two journals (Table 2), which assigns a substantially higher score to AFM.

The log-linear trend in citation accrual (Figure 3) appears generic (Burrell, 2007), applies to many journals, and is neatly summarised by the *h*-index, since it reflects the gradient of this relationship. Fewer than *h* papers (where *h* is the *h*-index) depart from this trend (i.e., those at the top left of Fig. 3), and appear to reflect the fortunate juxtaposition of easy accessibility and a topical issue, rather than research quality *per se*. The pattern revealed in Figure 3 leads to the suggestion that a classification of journals based on the *h*-index provides a better indicator for the RQF than the JIF. Figure 3 implies that the median journal contribution will be cited about *h*/3 times, an estimate that (unlike the JIF) is unaffected by the few papers that are frequently cited. A further advantage is that it can be calculated quickly and easily (e.g., with the PoP software; Harzing, 2007) for all journals, including those not recognised by Thomson Scientific. Figure 2 includes 43 journals recognised by Thomson Scientific, but also includes 43 journals with $h \geq 4$ not recognised by Thomson Scientific and without a JIF.

**Table 4**. Papers contributing to the PoP *h*-index, but excluded from the WoS *h*-index (2000-2007) for *Forest Ecology and Management.*

| Cites | Authors | Title | Year |
|---|---|---|---|
| 114 | de Vries et al | Intensive monitoring of forest ecosystems in Europe … | 2003 |
| 97 | Guariguata, Ostertag | Neotropical secondary forest succession … | 2001 |
| 72 | Marcot et al | Using Bayesian belief networks to evaluate fish and wildlife … | 2001 |
| 61 | Swank et al | Long-term hydrologic and water quality responses … | 2001 |
| 58 | Schoenholtz et al | A review of chemical and physical properties as indicators of forest soil … | 2000 |
| 56 | Ripple, Beschta | Wolf reintroduction, predation risk, and cottonwood recovery … | 2003 |
| 54 | Gardiner, Quine | Management of forests to reduce the risk of abiotic damage … | 2000 |
| 52 | Tiedemann et al | Solution of forest health problems with prescribed fire … | 2000 |
| 51 | Vesterdal et al | Change in soil organic carbon following afforestation … | 2002 |
| 51 | Griffis et al | Understory response to management treatments in northern Arizona … | 2001 |
| 48 | Liski et al | Increasing carbon stocks in the forest soils of western Europe. | 2002 |
| 48 | Knoepp et al | Biological indices of soil quality: an ecosystem case study of their use. | 2000 |
| 48 | Bowman et al | The association of small mammals with coarse woody debris … | 2000 |
| 47 | Fule et al | Comparing ecological restoration alternatives … | 2002 |
| 46 | Ketterings et al | Reducing uncertainty in the use of allometric biomass equations … | 2001 |
| 46 | Emborg et al | The structural dynamics of Suserup Skov … | 2000 |
| 45 | Pretzsch et al | The single tree-based stand simulator SILVA … | 2002 |
| 43 | Kavvadias et al | Litterfall, litter accumulation and litter decomposition rates … | 2001 |
| 43 | Yanai et al | Challenges of measuring forest floor organic matter dynamics … | 2000 |

Tables 2 and 3, and Figure 3 suggest that AFM and FEM are similar in many regards, but Figure 2 highlights the large discrepancy between the JIF and the *h*-index for these two journals. The total number of citations reported in Table 2 may shed some light on this difference. AFM appears to service a specialised audience that is more visible to Thomson Scientific than to Google Scholar. In contrast, FEM is cited in a substantial number of non-academic publications visible to Google Scholar, which reports 20% more citations than WoS (Table 2). An analysis of the differences in citation patterns for these 20,000 citations is a formidable task, but an insight may be gained by examining the differences in the few papers that contribute to the *h*-index estimated from TS and PoP records. The FEM papers contributing to the TS *h*-index (2000-2007) of 36 are not a complete subset of those contributing to the PoP *h*-index of 43, so there are 19 papers contributing to the PoP *h*-index but not the WoS *h*-index (Table 3). These 19 papers were cited a total of 1022 times, half of which (according to GS) accrued from WoS-listed journals, and the remainder from various sources including academic and government publications (Table 4). In the case of these 19 papers, there are at least as many citations from non-WoS sources as there are from WoS-listed journals. In this particular example, most these citations appear to *bona fide* and draw upon, rather than criticise the cited works. The citation of these FEM papers in academic theses and government reports (Table 5) suggests that FEM reaches practitioners as well as researchers. Although unproven, the difference in ratio of PoP:WoS h-indices (0.94 for AFM and 1.2 for FEM) seems to suggest that AFM is cited mainly by (and hence likely to be used mainly by) researchers, while the higher ratio for FEM may indicate greater uptake by practitioners.

**Table 5**. Sources of citations contributing to the PoP *h*-index but not to the WoS *h*-index (2000-2007) for *Forest Ecology and Management*.

| Source of citation | Cites (%) |
|---|---|
| WoS-listed journals (including FEM self-citations 9%) | 49 |
| Academic publications (including theses 10%) | 15 |
| Journals not listed by WoS (mostly refereed) | 12 |
| Government publications | 12 |
| Books | 6 |
| Conferences proceedings and presentations | 3 |
| Publications by NGOs and associations | 3 |
| Consultants reports and other commercial documents | 1 |
| Total | 100 |

## Discussion

There is no doubt that an *h*-index based on Google Scholar is imperfect (Jacso, 2008), in part because it can be manipulated with bogus documents on personal websites, and may be inflated by provocative contributions (such as A.D. Sokal's satirical 1996 contribution to *Social Text*, for which WoS records 18 citations, compared to 339 citations recorded by Google Scholar). However, the JIF is also imperfect, because it is available only for journals selected by Thomson Scientific, and because of limitations in the calculation of the JIF (Jacso, 2001; Dong et al., 2005; Vanclay, 2008).

The appendix offers a list of 180 forestry journals that have been cited at least once since 2000, and appear to contribute to forestry research and practice. This list has been compiled from the Thomson Scientific list, the Forest Science Database, Ulrich's Periodicals Directory, JournalSeek and Metla's Virtual Forestry library, further supplemented with Google Scholar searches for journals with a high frequency of forestry terms. The list was then culled to remove non-core forestry material, by removing titles that infrequently mentioned core forestry terms (such as forestry, silviculture, wood and timber). Google Scholar makes it easy to identify such journals efficiently, and to judge objectively whether or not a journal is central to a discipline. The list was ranked using *h*-indices computed by PoP (and for *Tree-Ring Research*, manually from GS data). RQF classifications (A1, A, B, C) were assigned to the 180 journals cited more than once during 2000-2007.

## Conclusion

The ranked list of journals provided in the appendix has several implications. Thomson Scientific may wish to recognise more of the high-ranked journals (such as *Dendrochronologia* with *h*-index 11), editors of some journals may wish to work with Google to make their contents more visible to search engines (e.g., *Ann. Forest Sci.* which is not recognised by Google as *Annals of Forest Science*, and *Tree-Ring Research* which does not provide the date of publication in Google-readable format), and editors of journals not published in English (which are disadvantaged in internet searches) may wish to add English abstracts and keywords to raise their profile.

Because of its broader coverage and despite known deficiencies, Hirsch's *h*-index based on Google Scholar data may be more useful than the Journal Impact Factor, as a measure of journal quality, and in providing a basis to rank journals.

**Appendix**. Ranked list of 180 selected forestry journals.

| Full Title | JIF | *h*-index 2000-7 | Class |
|---|---|---|---|
| *Forest Ecology and Management* | 1.839 | 43 | A1 |
| *Agricultural and Forest Meteorology* | 2.903 | 41 | A1 |
| *Journal of Vegetation Science* | 2.382 | 29 | A1 |
| *Tree Physiology* | 2.297 | 28 | A1 |
| *Plant Ecology (Vegetatio)* | 1.383 | 27 | A1 |
| *Canadian Journal of Forest Research* | 1.549 | 23 | A1 |
| *Forest Science* | 1.457 | 23 | A1 |
| *Journal of Forestry* | 1.188 | 23 | A1 |
| *Trees Structure and Function* | 1.461 | 22 | A1 |
| *International Journal of Wildland Fire* | 1.679 | 21 | A1 |
| *Annals of Forest Science* | 1.290 | 19 | A |
| *Agroforestry Systems* | 0.921 | 19 | A |
| *Agricultural and Forest Entomology* | 1.473 | 18 | A |
| *Scandinavian Journal of Forest Research* | 0.868 | 18 | A |
| *Holzforschung* | 1.014 | 17 | A |
| *Forest Policy and Economics* | 0.907 | 17 | A |
| *Forestry* | 0.847 | 16 | A |
| *Holz als Roh- und Werkstoff* | 0.514 | 15 | A |
| *Applied Vegetation Science* | 1.214 | 14 | A |
| *Silva Fennica* | 0.878 | 14 | A |
| *Forest Products Journal* | 0.387 | 14 | A |
| *Forestry Chronicle* | 0.831 | 13 | A |
| *Wood Science and Technology* | 0.740 | 13 | A |
| *International Forestry Review* | 0.618 | 12 | A |
| *Journal of Wood Science* | 0.574 | 12 | A |
| *Forest Pathology* | 0.729 | 11 | A |
| *New Forests* | 0.681 | 11 | A |
| *Dendrochronologia* | | 11 | A |
| *Unasylva* | | 11 | A |
| *Wood and Fiber Science* | 0.540 | 10 | A |
| *Revista Arvore* | | 10 | A |
| *Journal of Wood Chemistry and Technology* | 1.000 | 9 | B |
| *Southern Journal of Applied Forestry* | 0.704 | 9 | B |
| *Tree-Ring Research* | 0.625 | 9 | B |
| *Silvae Genetica* | 0.311 | 9 | B |
| *European Journal of Forest Pathology* | | 9 | B |
| *Journal of Forest Economics* | | 9 | B |
| *IAWA Journal* | 0.667 | 8 | B |
| *Western Journal of Applied Forestry* | 0.515 | 8 | B |
| *Forests, Trees and Livelihoods* | | 8 | B |
| *Forstwissenschaftliches Centralblatt (German Journal of Forest Science)* | | 8 | B |
| *Urban Forestry & Urban Greening* | | 8 | B |
| *Nordic Pulp & Paper Research Journal* | 0.562 | 7 | B |
| *Appita Journal* | 0.301 | 7 | B |
| *Journal of Tropical Forest Science* | 0.160 | 7 | B |
| *Australian Forestry* | | 7 | B |
| *Forest Genetics* | | 7 | B |
| *Journal of Sustainable Forestry* | | 7 | B |
| *Linye Kexue (Scientia Silvae Sinicae)* | | 7 | B |
| *Small-Scale Forestry* | | 7 | B |
| *Tasforests* | | 7 | B |
| *Northern Journal of Applied Forestry* | 0.779 | 6 | B |
| *European Journal of Forest Research* | 0.776 | 6 | B |
| *Allgemeine Forst- und Jagdzeitung* | 0.315 | 6 | B |
| *Ciencia Florestal* | | 6 | B |
| *Forst und Holz* | | 6 | B |
| *International Journal of Forest Engineering* | | 6 | B |
| *Investigacion Agraria. Sistemas y Recursos Forestales* | | 6 | B |
| *Journal of Forest and Livelihood* | | 6 | B |
| *Journal of Forest Research* | | 6 | B |
| *Scientia Forestalis* | | 6 | B |
| *Mokuzai Gakkaishi (Journal of the Japan Wood Research Society)* | 0.168 | 5 | B |
| *Bois et Forets des Tropiques* | | 5 | B |
| *Cerne* | | 5 | B |
| *Dendrobiology* | | 5 | B |
| *Floresta e Ambiente* | | 5 | B |
| *Forest Snow and Landscape Research* | | 5 | B |
| *Journal of Beijing Forestry University* | | 5 | B |
| *Journal of Forest Science* | | 5 | B |
| *Journal of the Japanese Forestry Society* | | 5 | B |
| *L'italia Forestale e Montana* | | 5 | B |
| *Revue Forestiere Francaise* | | 5 | B |
| *American Forests* | | 4 | C |
| *Baltic Forestry* | | 4 | C |
| *Floresta* | | 4 | C |
| *Forstarchiv* | | 4 | C |
| *Indian Forester* | | 4 | C |
| *Journal of Forest Planning* | | 4 | C |
| *Journal of Nanjing Forestry University* | | 4 | C |
| *New Zealand Journal of Forestry Science* | | 4 | C |
| *Quarterly Journal of Forestry* | | 4 | C |
| *Sherwood - Foreste ed Alberi Oggi* | | 4 | C |
| *Silva Lusitana* | | 4 | C |
| *Southern Hemisphere Forestry Journal* | | 4 | C |
| *Allgemeine Forst Zeitschrift* | | 3 | C |
| *Centralblatt fur das Gesamte Forstwesen* | | 3 | C |
| *Fire Ecology* | | 3 | C |
| *Forest Biometry Modelling and Information Sciences* | | 3 | C |
| *Forest Genetic Resources* | | 3 | C |
| *ITTO Tropical Forest Update* | | 3 | C |
| *Journal of Forest Engineering* | | 3 | C |
| *Journal of Fujian College of Forestry* | | 3 | C |
| *Journal of Northeast Forestry University* | | 3 | C |
| *Journal of the Institute of Wood Science* | | 3 | C |
| *Journal of the Japanese Forest Society* | | 3 | C |
| *Journal of the Korean Forestry Society* | | 3 | C |
| *Journal of Zhejiang Forestry College* | | 3 | C |
| *Madera y Bosques* | | 3 | C |
| *New Zealand Journal of Forestry* | | 3 | C |
| *Scottish Forestry* | | 3 | C |
| *Skoven* | | 3 | C |
| *Sylwan: Czasopismo Lesne* | | 3 | C |
| *Taiwan Journal of Forest Science* | | 3 | C |
| *World Forestry Research* | | 3 | C |

Forestry journals with 1-2 citations during 2000-07: *Agroforestry Today, Annales de la Recherche Forestiere au Maroc, Annali - Accademia Italiana di Scienze Forestali, Annals of Forestry, Australian Forest Grower, Austrian Journal of Forest Science, Drvna Industrija, East African Agricultural and Forestry Journal, Eurasian Journal of Forest Research, Fakta Skog, Folia Amazonica, Folia Forestalia, Folia Forestalia Polonica, Folia Oecologica, Forest and Bird, Forest and Landscape Research, Forest History, Forest History Today, Forest Inventory and Planning, Forest Pest and Disease, Forest Science and Technology, Forestry & British Timber, Forestry and Society, Forstzeitung, Frontiers of Forestry in China, Ghana Journal of Forestry, Holztechnologie, Indian Journal of Agroforestry, Indian Journal of Forestry, International Journal of Forest Usufructs Management, Iranian Journal of Forest and Poplar Research, Iranian Journal of Rangelands and Forests Plant Breeding and Genetic Research, Irish Forestry, Journal of Agriculture and Forestry, Journal of Forest Policy, Journal of Forest Products Business Research, Journal of Jiangsu Forestry Science & Technology, Journal of Research Forest of Kangwon National University, Journal of the Experimental Forest of National Taiwan University, Journal of The Fujian Agriculture and Forestry University, Journal of The Timber Development Association of India, Journal of Tropical Forest Products, Journal of Tropical Forest Resources, Journal of Tropical Forestry, Journal of Zhejiang Forestry Science and Technology, KFRI Journal of Forest Science (Seoul), Malaysian Forester, Metsätieteen Aikakauskirja, Myforest, Nederlands Bosbouw Tijdschrift, New Zealand Forestry, Nigerian Journal of Forestry, Norsk Skogbruk, Osterreichische Forstzeitung, Pakistan Journal of Forestry, PNG Journal of Agriculture Forestry and Fisheries, Protection Forest Science and Technology, Quarterly Journal of Chinese Forestry, Range Management and Agroforestry, Revista Chapingo: Serie Ciencias Forestales y del Ambiente, Revista Forestal Centroamericana, Revista Forestal Latinoamericana, Revista Forestal Venezolana, Revista Padurilor, Scandinavian Forest Economics, Schweizerische Zeitschrift Für Forstwesen, Temperate Agroforester, Thai Forest Bulletin, The Lao Journal of Agriculture and Forestry, Tohoku Journal of Forest Science, Tree-Ring Bulletin, Tropical Forestry, Turkish Journal of Agriculture and Forestry , Wood & Wood Products, Wood Research,* and *Wood Technology (Traeteknik).*